\documentstyle[preprint,aps,eqsecnum,epsf]{revtex}

\def\gtsim{$\raisebox{0.6ex}{$>$}\!\!\!\!\!\raisebox{-0.6ex}{$\sim$}\,\,$}
\def\ltsim{$\raisebox{0.6ex}{$<$}\!\!\!\!\!\raisebox{-0.6ex}{$\sim$}\,\,$}
\def\gtop{$\raisebox{0.9ex}{$g$}\!\!\!\!\!\raisebox{-0.2ex}{$\to$}\,\,$}

\begin{document}
\draft
\tightenlines

\title {\null\vspace*{-.0cm}\hfill {\small nucl-th/0112064} \\ \vskip
0.8cm
Dissociation of Heavy Quarkonia in the Quark-Gluon Plasma 
}

\author{Cheuk-Yin Wong}

\address{Physics Division, Oak Ridge National Laboratory, Oak Ridge,
TN 37831 USA}

\maketitle

\begin{abstract}
Using a temperature-dependent potential obtained from lattice gauge
calculations of Karsch $et~al.$, we study the stability of heavy
quarkonia in the quark-gluon plasma.  We find that only the
$\Upsilon(1S)$ and the $\eta_b(1S)$ are bound in the quark-gluon
plasma, and have a small binding energy.  The quark-gluon plasma may
be revealed by an $\Upsilon(1S)$ dilepton peak with an invariant mass
close to twice the current $b$ quark mass, which is lower than the
$\Upsilon(1S)$ mass in free space.  The quarkonia $\Upsilon(1S)$ and
$\eta_b(1S)$ can dissociate by collision with quarks and gluons in the
quark-gluon plasma.  The $\Upsilon(1S)$ and the $\eta_b(1S)$ can also
dissociate spontaneously at temperatures above the dissociation
temperature $1.11 T_c$, where $T_c$ is the quark-gluon plasma phase
transition temperature.  At temperatures slightly above the
dissociation temperature these states appear as resonances, which
provides another signature for the quark-gluon plasma.

\end{abstract}

\newpage

\section{Introduction}

The suppression of heavy quarkonium production in a quark-gluon plasma
has been a subject of intense interest since the pioneering work of
Matsui and Satz \cite{Mat86}.  Initial insight into the dissociation
temperatures of heavy quarkonium was further provided by Karsch, Mehr,
and Satz \cite{Kar88}.  Recently, using the temperature-dependent
$Q$-$\bar Q$ interaction inferred from lattice gauge calculations of
Karsch $et~al.$ \cite{Kar00}, Digal, Petreczky, and Satz
\cite{Dig01a,Dig01b} reported theoretical results for the dissociation
temperatures of heavy quarkonia in hadronic matter $(T<T_c)$ and in
the quark-gluon plasma $(T\ge T_c)$, where $T_c$ is the quark-gluon
plasma phase transition temperature.  Subsequent analysis of the
dissociation temperatures of heavy quarkonia at $T<T_c$ using
different interactions and selection rules was given in detail in
\cite{Won01a} and summarized in \cite{Won01b}.  The dissociation of
heavy quarkonia by thermalization and by collisions with particles in
the medium has also been investigated
\cite{Won01a,Won01b,Won00,Won00a,Bar00,Won01,Kha94,Mat98,Hag00,Hag01,Lin99,Lin00,Lin01,Oh00}.

For $T>T_c$, the equilibrium state of matter is the quark-gluon
plasma.  If the interaction between a $Q\bar Q$ and the QGP is weak,
the $Q\bar Q$ system can be color-singlet or color-octet.  Using
perturbation theory as a guide, Digal $et~al.$ \cite{Dig01b} extracted
the color-singlet potential $V_1(r,T)$ and color-octet potential
$V_8(r,T)$ of a heavy quark pair from the Polyakov loop $\langle
L(0)L^\dagger (r)\rangle$.  They used the relation \cite{Mcl81}
\begin{eqnarray}
\label{eq:loop}
 {\langle L(0)L^\dagger (r)\rangle \over  \langle L^2 \rangle}   
={1\over 9} \exp \left [{-V_1(r,T) \over T } \right ] 
+{8 \over 9}  \exp \left [ {-V_8(r,T) \over T } \right ]
\end{eqnarray}
where $\langle L^2 \rangle$ denotes $\langle L(0)L^\dagger (r)\rangle$
at $r\to \infty$.  This relation is valid when the $Q\bar Q$ interacts
weakly with a perturbative color medium so that color-singlet and
color-octet $Q\bar Q$ states are approximate eigenstates of the
system.  They also made the assumptions
\begin{eqnarray}
V_1(r,T)=-{4\over 3}
{\alpha(T)\over r} e^ {-\mu(T) r},
\end{eqnarray}
and
\begin{eqnarray}
\label{eq:per}
V_8(r,T)={1\over 6} c(T) {\alpha(T)\over r} e^ {-\mu(T) r},
\end{eqnarray}
as suggested by perturbation theory (for which $C(T)=1$).  They
assumed that $\mu(T)$ was $1.15 T$ from the systematics at high
temperatures and chose functions $\alpha(T)$ and $c(T)$ to fit the
Polyakov loop results of Karsch $et~al.$ \cite{Kar00}.  They examined
color-singlet heavy quarkonia in the quark-gluon plasma, even for
temperatures slightly greater than $T_c$ \cite{Dig01b}.

We shall restrict our interest to the region of temperatures slightly
greater than $T_c$, which is the most important region for the
investigation of the stability of heavy quarkonia in the quark-gluon
plasma.  In this temperature region, the screening effects are
non-perturbative.  Evidence for non-perturbative screening comes from
numerical lattice gauge calculations of Karsch et al. \cite{Kar01a}
who show that the Debye screening mass is significantly larger than
the leading-order perturbative value of $m_D=gT\sqrt{1+n_f/6}$ and
that the screening mass remains about three times larger than the
perturbative Debye mass $m_D$ even up to temperatures as high as
100$T_c$.  Similarly, the $rT$ dependence as shown in Eqs.\
(\ref{eq:loop})-(\ref{eq:per}) with the perturbative value of $C(T)=
1$ is approximately valid only for $T>3T_c$ \cite{Kar01a}.

For the region of temperatures close to $T_c$ in which we shall be
interested, $C(T)$ obtained from the procedures of Ref.\ \cite{Dig01b}
is about 0.1 which deviates significantly from the perturbative value
of $C(T)=1$, and the system cannot be described as a weakly coupled and
perturbatively screened system \cite{Kar01a}.  Non-perturbative
screening effects are important and the extrapolation procedure in the
non-perturbative region employed in Ref.\ \cite{Dig01b} is
ambiguous. Specifically, the Polyakov loop provides a single number
for a given $r$ and $T$, but the above separation procedure yields
quantitative determination for the physical quantities of $V_1$,
$V_8$, and the (dynamical) relative weight of the color-singlet and
color-octet components.  Many assumptions are needed to extract these
three physical quantities out of a single quantity of the Polyakov
loop.  Without theoretical non-perturbative results to guide us, these
additional assumptions are not unique.  Different assumptions will
lead to different extracted quantities in the non-perturbative region.

Related to the question of the separation of the potential is the
question whether purely color-singlet and color-octet $Q\bar Q$ states
are approximate eigenstates in the non-perturbative region of
temperatures in QGP, when $T$ is just slightly greater than $T_c$.
Purely color-singlet and color-octet $Q\bar Q$ states are approximate
eigenstates of the system in the high-temperature, perturbative region
for which Eq.\ (\ref{eq:loop}) is valid.  This however cannot be the
case in the non-perturbative region of $T$\gtsim$T_C$.  A constituent
of a $Q\bar Q$ in a deconfined medium can interact with a constituent
of the medium ($q,\bar q,$ and $g$) by exchanging a single gluon.
Such a gluon exchange alters the color of the $Q\bar Q$ from a
color-singlet state to a color-octet state and vice versa.  The color
of the medium will undergo corresponding changes so that the color of
the whole system is neutral.  Such a color-changing interaction mixes
the color states of the heavy quarkonium.  We shall show below in
Section II that the degree of mixing increases with the magnitude of
the ratio of the color-changing interaction matrix element $\lambda$
to the energy difference $\Delta$ between the unperturbed color-octet
and color-singlet states.  In the non-perturbative region of $T$ in
the QGP, $|\lambda/\Delta|$ is not small and the eigenstates of the
system become states of mixed colors.  A purely color-singlet or purely
color-octet state cannot be a stationary state of the non-perturbative
system.

Even though a $Q\bar Q$ state in the non-perturbative region of
temperatures cannot be separated into eigenstates of pure color, it is
still a meaningful question to find out how the $Q$ interacts with the
$\bar Q$ in such a system, whatever the color admixture of the
resultant $Q\bar Q$ may be.  In lattice gauge calculations the
expectation value of the Polyakov loop operator is evaluated allowing
all possible configurations of the gauge fields and fermion fields at
all lattice points and links, and this corresponds to all color states
of the $Q\bar Q$ and the medium.  The gauge field variables along the
path of the Polyakov loop provide information on the evolution of the
heavy-quark system.  The greatest contributions to the Polyakov loop
at a given $Q$-$\bar Q$ separation will come from those gauge fields
and quark fields which adjust themselves in magnitude and in color
admixture to give states of lowest energies of a $Q$-$\bar Q$ in the
medium.  As the eigenstates of the system are states with mixed colors
(see Section II), the Polyakov loop provides information on the free
energy of mixed-color states.  Carrying this calculations for
different separations between the $Q$ and the $\bar Q$, one obtains
the free energy as a function of the separation, which provides the
spatial variation of the effective interaction between the $Q$ and the
$\bar Q$.  It can be used to find out if the interaction is strong
enough to bind the $Q$ with the $\bar Q$ in the non-perturbative
deconfined medium.  Therefore, in this non-perturbative medium with
strong interactions, the effective interaction potential $V_{Q\bar
Q}(r,T)$ between the $Q$ and the $\bar Q$, is related to the
expectation value of the Polyakov loop from lattice gauge theory by
\begin{eqnarray}
\label{eq:pol}
{\langle L(0)L^\dagger (r)\rangle \over \langle L^2\rangle}
= \exp \left [ {-V_{Q\bar Q}(r,T) \over T} \right ].
\end{eqnarray}
The quantity $V_{Q\bar Q}(r,T)$ was extracted previously from the
lattice gauge calculations by Karsch $et~al.$ \cite{Kar00}.  The
binding energies of quarkonia as a function of the temperature will
provide useful plasma diagnostic information on the dissociation
temperatures and quarkonia masses in the nonperturbative quark-gluon
plasma phase.

As a heavy quarkonium becomes unbound, it appears as a resonance in
the continuum.  Such a resonance occurs only in the quark-gluon plasma
at a certain range of temperatures and may be used as a signature of
the quark-gluon plasma \cite{Won97}.  Following the method used
previously by Calogero\cite{Cal67} and by Wong and Chatterjee
\cite{Won97}, we shall study the resonance structure of heavy
quarkonia in the quark-gluon plasma.

Quarkonia can also dissociate by collision with particles in the
medium
\cite{Won01a,Won01b,Won00,Won00a,Bar00,Won01,Kha94,Mat98,Hag00,Hag01,Lin99,Lin00,Lin01,Oh00}.
In particular, we will study $\Upsilon(1S)+g \to b+\bar b $.  Such a
dissociation process is similar to the photodissociation of a deuteron
after absorbing a photon.  This dissociation process can be studied by
using a multipole expansion of the gluon field, analogous to the
electromagnetic case discussed in detail by Blatt and Weisskopf
\cite{Bla52}.  As the gluon dissociation cross section has also been
obtained previously by Peskin and Bhanot \cite{Pes79,Bha79} and
subsequently used by Kharzeev and Satz \cite{Kha94}, we show in the
Appendix the equality of the cross sections obtained by the two
formalisms for the simple case considered by Peskin and Bhanot.
 
In Section II, we discuss the color structure of a $Q\bar Q$ and its
dependence on the interaction between the constituents of the $Q \bar
Q$ and the constituents of the medium.  In Section III, we describe
the Schr\" odinger equation used to calculate the energies and wave
functions of quarkonium states.  We introduce an analytical form of
the effective screening potential $V_{Q\bar Q}$ to represent the
results of lattice gauge calculations.  The quarkonium energies and
the dissociation temperatures are calculated in Section IV. In Section
V, we consider continuum states and follow the resonances as functions
of the quark-gluon plasma temperature.  In Section VI we evaluate the
cross section of the dissociation process $\Upsilon(1S)+g \to b+\bar b
$.  A discussion of our results is given in Section VII.

\section{ Color Structure of a $Q\bar Q$ in the QGP
}

The color structure of a $Q\bar Q$ placed in a deconfined medium
depends on the interaction between the constituents of the $Q\bar Q$
and the constituents of the deconfined medium.  We can consider first
two extreme cases of very weak and very strong interactions as
illustrated in Fig.\ 1.

If the interaction between the $Q$ and the $\bar Q$ of the quarkonium
with $q,\bar q,$ and $g$ is weak as in a perturbative medium, [Fig.\
1(a)], the $Q$ and the $\bar Q$ can combined together to form pure
color-singlet and color-octet eigenstates. The concept of a purely
color-singlet or color-octet state is useful as an approximate
description, when the interaction between the $Q$ and the $\bar Q$
with the quark matter particles can be ignored.

\vspace*{3.7cm}
\epsfxsize=300pt
\includegraphics{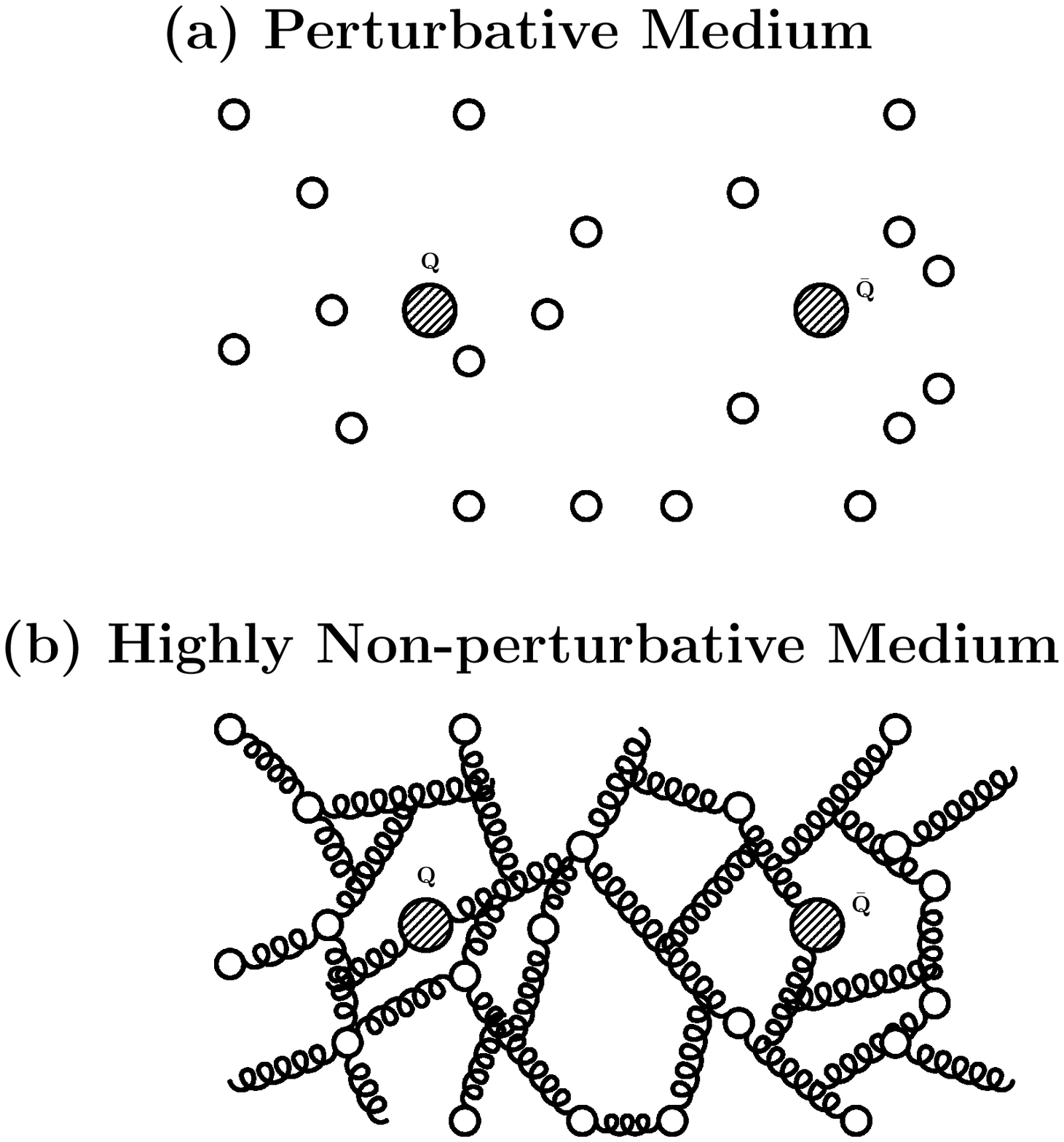}
\vspace*{+8.3cm}\hspace*{1cm}
\begin{minipage}[t]{12cm}
\noindent {\bf Fig.\ 1}.  {$Q$ and $\bar Q$ in a medium with
perturbatively weak interactions (a), and in a highly non-perturbative
medium with strong interactions (b).  Open circles represent quarks,
antiquarks, and gluons in the plasma.}
\end{minipage}
\vskip 4truemm
\noindent

In the other extreme of very strong interactions as illustrated in
Fig.\ 1(b), the $Q$ and the $\bar Q$ are intimately linked to the
medium particles by gluons.  The isolation of the $Q\bar Q$ system is
impossible and one cannot speak unambiguously of a purely
color-singlet or color-octet state.

We discuss how we can pass from one limit to the other limit with a
schematic model.  We consider a $Q \bar Q$ system in a medium and use
$(Q\bar Q)^{(c)}$ basis states where $c=0$ is for color-singlet and
$c=8$ for color octet.  We consider for simplicity the medium
represented by $M^{(c)}$ where $c$ is again the color index.  As the
color of the whole system consisting of both $(Q\bar Q)^{c}$ and
$M^{c}$ must be color neutral, a state of the system is described by
\begin{eqnarray}
|\psi\rangle = a_0 (Q\bar Q)^{(0)} M^{(0)} + a_8 [(Q\bar Q)^{(8)}
 M^{(8)}]^{(0)}
\end{eqnarray}
One can represent this admixture by a column vector with elements
$a_0$ and $a_8$.  In the absence of the interaction between the $Q$
and the $\bar Q$ with the medium particles, the color-singlet state
and the color-octet state can be separated and we denote the energy
difference between the unperturbed color-octet state and the
color-singlet state by $\Delta=E^{(8)}-E^{(0)}$.

By the interaction $v_{Aa}(A+a ~~$ \gtop $~~ A' +a')$ in which the
constituent $A$ in $(Q\bar Q)^{(c)}$ exchange a gluon $g$ with the
constituent $a$ in $M^{(c)}$ to become $A'$ and $a'$, the color of
both $(Q\bar Q)^{(c)}$ and $M^{(c)}$ will change.  We define the
color-changing transition matrix element $\lambda$ as
\begin{eqnarray}
\lambda=\langle [(Q\bar Q)^{(8)} M^{(8)}]^{(0)} ~|v_{Aa}|~(Q\bar Q)^{(0)}
M^{(0)}\rangle
\end{eqnarray}
The Hamiltonian for such a system is
\begin{eqnarray}
H=\left ( \matrix { -\Delta/2   & \lambda \cr
                    \lambda     & \Delta/2 \cr} \right ).
\end{eqnarray}
The lowest energy eigenstate of the system is then
\begin{eqnarray}
\label{eq:eig0}
|\psi\rangle_0 =  { 1+(1+4\lambda^2/\Delta)^2)^{1/2} \choose
-2\lambda/\Delta}
{1 \over \sqrt{2}[1+4\lambda^2/\Delta^2 
+ (1+4\lambda^2/\Delta^2)^{1/2}]^{1/2}}.
\end{eqnarray}
The excited eigenstate is
\begin{eqnarray}
|\psi\rangle_1 =  { 2\lambda/\Delta \choose
1+(1+4\lambda^2/\Delta)^2)^{1/2}
}
{1 \over \sqrt{2}[1+4\lambda^2/\Delta^2 
+ (1+4\lambda^2/\Delta^2)^{1/2}]^{1/2}}
\end{eqnarray}

Thus, the color-structure of the system depends on the ratio
$|\lambda/\Delta|$.  In the confined phase, $\Delta$ is very large,
and so $|\lambda/\Delta|$ is very small; the lowest eigenstate is
purely color singlet.  In the perturbative deconfined phase at very
high temperatures, $\lambda$ is small and $|\lambda/\Delta|$ is also
small; the eigenstates are nearly pure in color.

On the other hand, if $|\lambda/\Delta|$ is large, the eigenstates of
the system is an admixture of color-singlet and octet state.  The
nonperturbative region of temperature is a case of strong interactions
between the $Q$ (or $\bar Q$) and the medium particles.  We can
estimate the ratio $|\lambda/ \Delta|$ which determines the color
admixture.  The energy difference $\Delta$ between color-octet and
color-singlet states is approximately the dissociation energy of the
color-singlet state.  Earlier estimates of the dissociation energy of
$J/\psi$ in the hot hadron phase show that it decreases with
increasing temperatures, resulting in zero dissociation energy
\cite{Won01a} or nearly zero dissociation energy (see Fig.\ 3 of
\cite{Dig01a}) for $J/\psi$ at $T$\ltsim$T_c$.  For $T$\gtsim$T_c$,
screening of deconfined quark matter will tend to weaken further the
$c$-$\bar c$ attraction.  (Calculations in Section IV is consistent
with a vanishing value of the dissociation energy of $J/\psi$.)  Thus,
$\Delta$ is zero or nearly so for $c\bar c$.  For $b \bar b$, the
dissociation energy is about 0.15 GeV for $\Upsilon (1S)$
\cite{Won01a} at $T$\ltsim$T_c$. For $T$\gtsim$T_c$, screening in the
quark-gluon plasma makes the dissociation energy even smaller.  (In
Section VI, the dissociation energy is found to be of order 0.01 GeV.)
On the other hand, for a QGP at $T=200$ MeV, the average separation
between the quanta in the QGP is approximately $d$= 0.6 fm .  The
transition matrix element is of order $ \lambda\sim {\alpha_s }\exp
\{-m_{\rm th} d\}/d $ where the thermal mass $m_{\rm th}\sim gT$ and
$g^2=\alpha_s/4\pi$.  We find that for $\alpha_s = 0.2$, $\lambda$ is
of order 0.06 GeV.  The ratio $|\lambda/ \Delta|$ is also quite large
for $b\bar b$, if we take $\Delta\sim 0.01$ GeV.  Therefore, the
$Q\bar Q$ system is strongly mixed in color in the nonperturbative
deconfined region of temperatures slightly greater than $T_c$.

Even though a $Q\bar Q$ state in the non-perturbative region of
temperatures cannot be separated into eigenstates of pure color, it
remains a meaningful question to find out how the $Q$ interacts with
the $\bar Q$ in such a system.  As explained in the Introduction, the
expectation value of the Polyakov loop operator gives the free energy
of the system of the $Q\bar Q$ coupled to the medium in a mixed-color
state.  It provides the effective interaction potential $V_{Q\bar
Q}(r,T)$ between the $Q$ and the $\bar Q$ in in the region of $T$\gtsim$T_c$.
We shall use this interaction potential to examine the stability of
the $Q\bar Q$ system in the next few sections.

\section{ Schr\" odinger Equation for Heavy Quarkonium States}

The energy $\epsilon(T)$ of a heavy quarkonium state $(Q\bar
Q)_{JLS}$, measured relative to the mass of the $Q$ and the $\bar Q$,
can be obtained by solving for the associated eigenvalue of the Schr\"
odinger equation
\begin{eqnarray}
\label{eq:sch}
\biggl \{ - { \hbar^2 \over 2\mu_{Q\bar Q}}\nabla^2   
+ V_{Q\bar Q}(r,T) \biggr \} 
\psi_{JLS} (\bbox{r},T)
 = \epsilon(T) \psi_{JLS} (\bbox{r},T),
\end{eqnarray}
where $\mu_{Q\bar Q}=m_Q/2$ is the reduced mass.  The quarkonium is
bound if $\epsilon(T)$ is negative, and is unbound and dissociates
spontaneously into a $Q$ and a $\bar Q$ if $\epsilon(T)$ is positive.

The color charges of the $Q$ and the $\bar Q$ are screened in the
quark-gluon plasma.  We can conveniently represent $V_{Q\bar Q}(r,T)$
by a screened Yukawa potential with an effective strength $\alpha_{\rm
eff}(T)$ and a screening mass parameter $\mu(T)$.  We shall be
interested in temperatures slightly greater than $T_c$.  The $Q$-$\bar
Q$ interaction obtained from the lattice calculations of Karsch
$et~al.$ \cite{Kar00} for $1.2T_c>T>T_c$ in the quark-gluon plasma
phase can be approximated by
\begin{eqnarray}
\label{eq:pot}
V_{Q\bar Q}(r,T)=-{4\over 3}\alpha_{\rm eff}(T) {e^{-\mu (T) r}\over r}
\end{eqnarray}
where
\begin{eqnarray}
\label{eq:pot1}
&&\alpha_{\rm eff}(T)=0.20,  {~~~~~~~~} 
\mu(T)=\mu_0 + \mu_T (T/T_c-1),
\\
&&\mu_0=0.323 {\rm ~~GeV,~~~~~and~~~~~~} \mu_T=3.04 {\rm ~~~GeV}.
\nonumber
\end{eqnarray}

Fits to $V_{Q\bar Q}(r,T)$ using the above form are shown in Fig.\ 2.
We include the color-singlet factor $(-4/3)$ in the above definition
(\ref{eq:pot}) for $\alpha_{\rm eff}(T)$ in order to have a comparison
of the strength of the $Q$-$\bar Q$ interaction in the quark-gluon
plasma phase with the color-singlet $Q$-$\bar Q$ interaction seen at
$T<T_c$ \cite{Won01a}.  The strength parameter $\alpha_{\rm eff}(T)$
has very weak temperature dependence, and can be adequately
represented by the constant $\alpha_{\rm eff}(T)=0.20$.  This value of
$\alpha_{\rm eff}(T)=0.20$ in the quark-gluon plasma phase is smaller
than the value of $\alpha_s \sim 0.32$ for charmonium and $\alpha_s
\sim 0.24$ for bottomium in the screened color-Coulomb potential for
temperatures below $T_c$ \cite{Won01a}.  The screening mass parameter
$\mu_0$ for $T \ge T_c$ is greater than the screening mass parameter
of $\mu_0 = 0.28$ GeV used for $T<T_c$ in \cite{Won01a}.  One
concludes that because of the screening due to deconfined quarks and
gluons, the effective potential between the $Q$ and the $\bar Q$ in
the quark-gluon plasma remains attractive, but its strength is weaker
and its range is shorter than the color-Coulomb interaction seen in
hadronic matter.  The range of the screened potential decreases with
increasing temperatures.

\vspace*{3.7cm}
\epsfxsize=300pt
\includegraphics{vqgpvt0.eps}
\vspace*{+8.3cm}\hspace*{1cm}
\begin{minipage}[t]{12cm}
\noindent {\bf Fig.\ 2}.  {The effective potential $V_{Q\bar Q}(r,T)$
between the $Q$ and the $\bar Q$ in a quark-gluon plasma at various
temperatures.  The symbols represent the results of the lattice gauge
calculations of \cite{Kar00} reported in \cite{Dig01b}.  The curves in
the main figure are from the paramterization of Eqs.\ (\ref{eq:pot})
and (\ref{eq:pot1}).  The inserted figure shows the effective
potential at $T=0$(for charmonium) and $T=1.03T_c$.}
\end{minipage}
\vskip 4truemm
\noindent

\section{Bound states and dissociation temperature}

The eigenvalues of the Hamiltonian can be obtained by integrating the
Schr\" odinger equation numerically and requiring an exponentially
decaying wave function at large distances.  For our numerical
calculations we use a current $c$ quark mass $m_c=1.25$ GeV, and a
current $b$ quark mass $m_b=4.2$ GeV \cite{pdg00}.

We find that the potential of Eqs.\ (\ref{eq:pot}) and (\ref{eq:pot1}),
as determined by the lattice gauge calculations of Karsch $et~al.$, is
too shallow to hold a bound charmonium state.  Because of the large
$b$ quark mass, we can neglect the spin-orbit and hyperfine
interactions.  In this approximation, the $L=0$ $\Upsilon(1S)$ and
$\eta_b(1S)$ states are degenerate, as are the $L=1$ $\chi_b(1P)$ and
$h_b(1P)$ states.  For $b\bar b$ the $L=1$ $\chi_b(1P)$ and $h_b(1P)$
states are unbound, and the only bound states are the $L=0$
$\Upsilon(1S)$ and $\eta_b(1S)$ states.  The energies of the
$\Upsilon(1S)$ and $\eta_b(1S)$ states as functions of the quark-gluon
plasma temperature are shown in Fig.\ 3. They become less bound with
increasing temperature.  The dissociation temperature of
$\Upsilon(1S)$ and $\eta_b(1S)$ is $T_d=1.11T_c$, as determined by the
condition $\epsilon=0$.  The root-mean-squared radius of
$\Upsilon(1S)$ at $T=T_c$ is 0.80 fm, indicating a large radius
because of the weak binding.  The rms radius increases with
temperature.  At $T=1.10T_c$ which is close to the dissociation
temperature, the rms radius has increased to 3.3 fm.

\vspace*{2.8cm} \epsfxsize=300pt \includegraphics{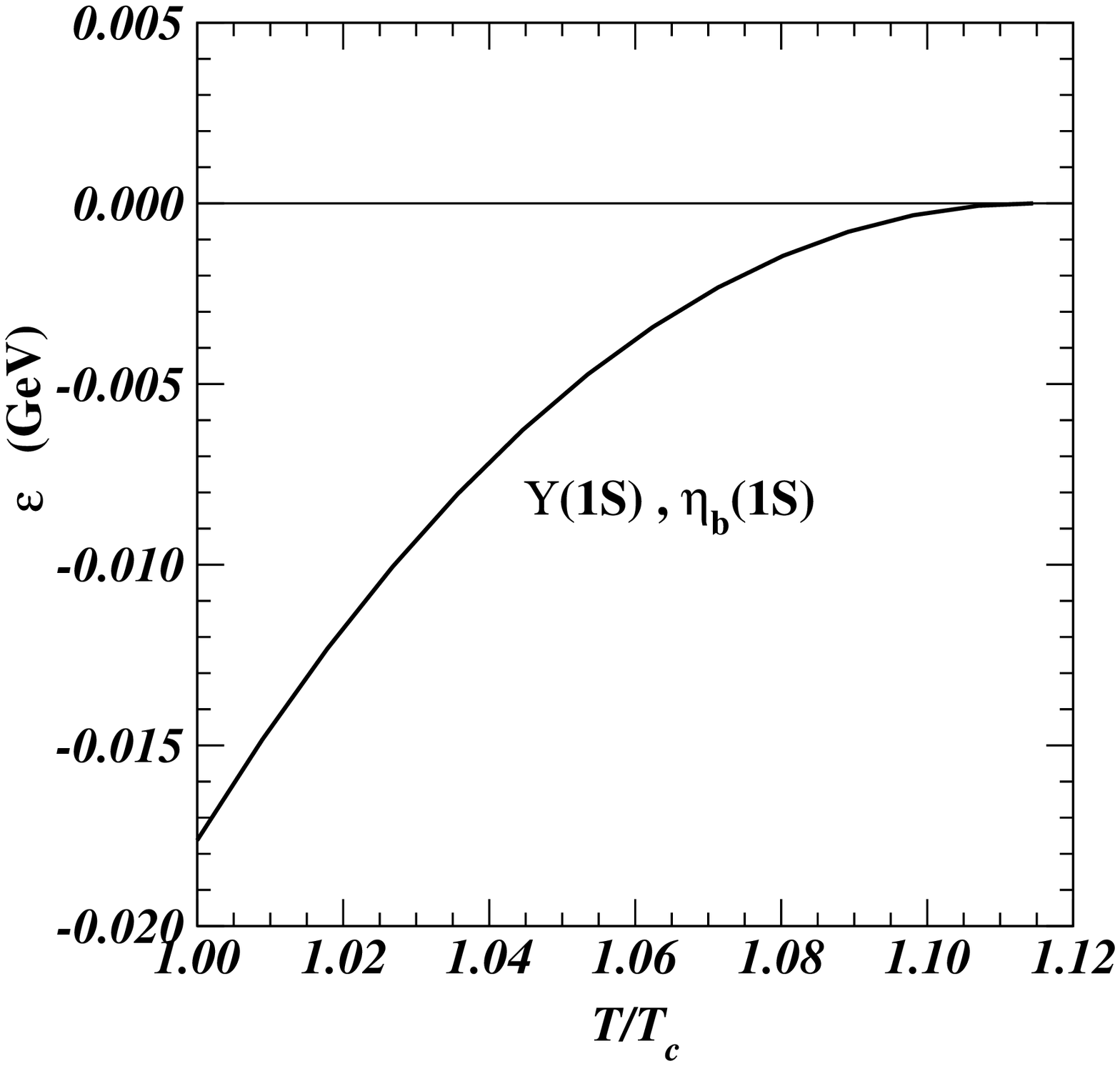}
\vspace*{+7.9cm}\hspace*{1cm}
\begin{minipage}[t]{12cm}
\noindent {\bf Fig.\ 3}.  {The energies of $\Upsilon(1s)$ and
$\eta_b(1S)$ states as a function of temperature.  }
\end{minipage}
\vskip 4truemm
\noindent 

Previously we found that the $\chi_{b1}$ and $\chi_{b2}$ states were
stable against dissociation for $T<T_c$.  In the quark-gluon plasma
phase, however, screening by the deconfined $q$, $\bar q$, and $g$
alters the interaction between $Q$ and $\bar Q$.  The interaction
does not need to be a continuous function of the temperature at $T_c$,
because a new type of screening arises from deconfined $q$, $\bar q$,
and $g$ in the quark-gluon plasma phase for $T\ge T_c$.  Consequently,
the quarkonium energies do not need to be continuous across
the QGP phase transition boundary.

For completeness we show the heavy quarkonium dissociation
temperatures in Table I where we also include those shown in Table III
of \cite{Won01a}.  The dissociation temperature of $T_d/T_c=1.00+$ for
$\chi_b$ and $h_b$ in Table I denotes $T_d/T_c=1.00$ in the
quark-gluon plasma phase.  The dissociation temperatures obtained here
are substantially lower than the values of 1.13$T_c$ for the
color-singlet $\chi_b$ and 2.31$T_c$ for the color-singlet
$\Upsilon(1S)$ obtained by Digal $et~al.$ \cite{Dig01b} using the
color-singlet potential $V_1(r,T)$.  These differences arise because
the present investigation assumes a $Q\bar Q$ pair that is immersed in
and interacting strongly with the color medium of the quark-gluon
plasma, and the work of Digal $et~al.$ \cite{Dig01b} considers the
dissociation of a color-singlet $Q\bar Q$ pair in the quark-gluon
plasma.  Hence there are significant differences in the dissociation
temperatures.

\vspace*{0.5cm} \centerline{Table I.  The dissociation temperatures
$T_d$ in units of $T_c$ for various heavy quarkonia}
{\vspace*{0.3cm}\hspace*{0.0cm}
\begin{tabular}{|c|c|c|c|c||c|c|c|c|c|c|} \hline
\vspace*{-0.0cm}
{\rm Heavy           }& $~~\psi'~~$ &$~~\chi_{c2}~~$ 
                      & $~~\chi_{c1}~~$ & $~~J/\psi~~$ 
                      & $~~\Upsilon''~$ & $~~\chi_{b2}'~$ 
                      & $~~\chi_{b1}'~~$
                      & $~~\Upsilon'~~$ 
                      & $\chi_{b},h_b$ & $\Upsilon,\eta_b$
                        \\
{\rm Quarkonium}     &  &  &  & 
		     &  &  &
                     &  &  &  \\
\hline
 $T_d/T_c$             & 0.50 & 0.91 & 0.90&  0.99
                   & 0.57 & 0.82  & 0.82  
                   & 0.96 & 1.00+ & 1.11   \\
\hline  
\vspace*{-0.0cm}
 $T_d/T_c$         & 0.1-0.2 & 0.74 & 0.74& 1.10
                   & 0.75 & 0.83  & 0.83  
                   & 1.10 & 1.13  & 2.31   \\
(Digal $et~al.$\cite{Dig01b})     &  &  &  & 
		     &  &  &
                     &  &  &  \\
\hline
\end{tabular}
}
\vskip 0.6cm

\section{Quarkonium states in the continuum}

To study the decay of a heavy quarkonium resonance, it is of interest
to examine the related $Q$-$\bar Q$ scattering process.  The
scattering phase shifts with the screened potential $V_{Q\bar Q}(r,T)$
will provide information regarding the locations and the widths of
heavy quarkonium resonances in the continuum.

To obtain the continuum wave function in the scattering problem, we
use the phase-angle method discussed in detail by Calogero
\cite{Cal67} and used previously by Wong and Chatterjee \cite{Won97}.
We write the wave function as
\begin{eqnarray}
\psi_{JLS}(\bbox{r}) =R_{JLS}(r)  {\cal Y}_{JLS}(\hat {\bbox{r}})
={u_{JLS}(r) \over kr} {\cal Y}_{JLS}(\hat {\bbox{r}})
 \sqrt {{4 \pi \over 2l+1}},
\end{eqnarray}
and represent it in terms of an amplitude
$\alpha_{L}(r)$ and a phase shift $\delta_{L}(r)$
\begin{eqnarray}
\label{eq:wf}
u_{JLS}(r)={\alpha_{L}(r) \over \alpha_{L}(\infty)} {\hat D}_{L}(kr) \sin ({{\hat
\delta}_{L} (kr) + \delta_{L}(r)}),
\end{eqnarray}
with the boundary condition that $\delta_{L}(r\rightarrow 0)=0.$ We
have normalized the radial wave function such that it coincides with
the normalization of Eq. (4.32) of Blatt and Weisskopf \cite{Bla52}.
In the case of a free particle, $u_{JLS}(r) \sim krj_L(kr)$.  The
functions ${\hat D}(kr)$ and ${\hat \delta}(kr)$ are known functions
determined from the free particle wave functions \cite{Cal67},
\begin{eqnarray}
\label{eq:dhat}
{\hat D}_0(x)=1, ~~~{\hat D}_1(x)=(1+1/x^2)^{1/2},~~
\end{eqnarray}
and
\begin{eqnarray}
\label{eq:delh}
{\hat \delta}_0(x)=x, ~~~{\hat \delta}_1(x)=x-\tan^{-1}x.
\end{eqnarray}
The phase shift function $\delta_{L}(r)$ depends on the interaction
$U(r)=2\mu_{Q\bar Q}V_{Q\bar Q}(r)$.  The equation for $\delta_{L}(r)$
is given by \cite{Cal67}
\begin{eqnarray}
\label{eq:edelta}
{d \over dr} \delta_{L}(r)=
- {U(r) \over k} {\hat D}_{L}^2(kr)
\left \{ \sin[{\hat \delta}_{L}(kr)+\delta_{L}(r)]\right \}^2.
\end{eqnarray}
After the function $\delta_{L}(r)$ is evaluated, the amplitude
$\alpha_L(r)$ can be obtained from $\delta_{L}(r)$ by 
\begin{eqnarray}
\label{eq:alp}
\alpha_{L}(r)=\exp \left \{{1 \over 2k}\int_0^r ds\, U(s)\,
{\hat D}_{L}^2(ks) \, \sin 2[{\hat
\delta}_{L}(ks)+\delta_{L}(s)] \right  \}.
\end{eqnarray}

\vspace*{1.5cm} \epsfxsize=300pt 
\includegraphics{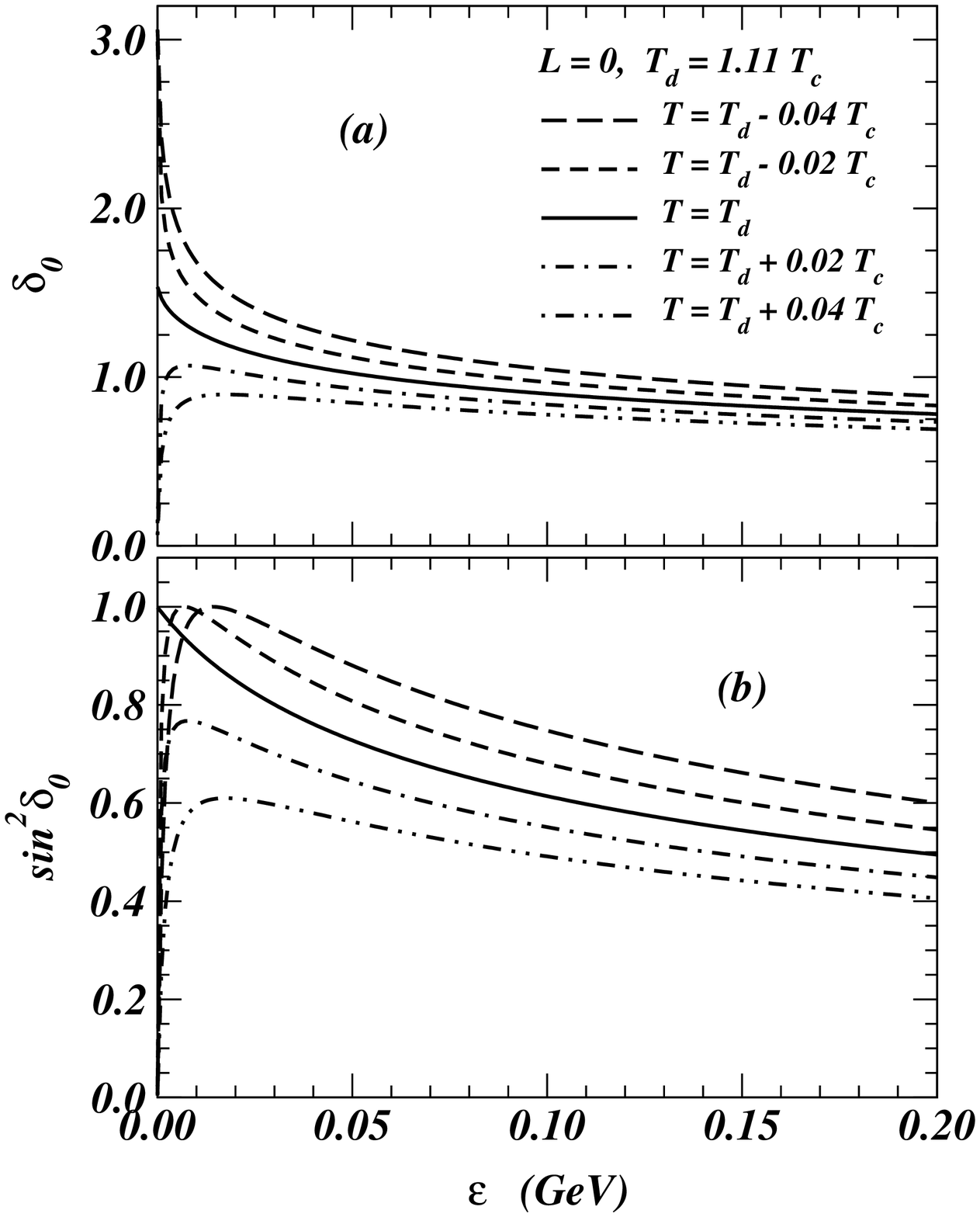}
\vspace*{+12.0cm}\hspace*{1cm}
\begin{minipage}[t]{12cm}
\noindent { {\bf Fig.\ 4.} ($a$) The phase shift $\delta_0$ and ($b$)
$\sin ^2 \delta_0$ for the $L=0$ scattering of $Q$ and $\bar Q$ as a
function of energy $\epsilon$ for different $T$ values near the
dissociation temperature $T_d$.}
\end{minipage}
\vskip 4truemm
\noindent 

Numerical integration of Eq.\ (\ref{eq:edelta}) gives the asymptotic
phase shift $\delta_L\equiv \delta_L (\infty)$ and the continuum wave
function.  We show the phase shift $\delta_0$ in Fig.\ 4($a$) and
$\sin^2 \delta_0$ in Fig.\ 4($b$) for the $L=0$ $b\bar b$ state as a
function of energy $\epsilon$ at various temperatures.  The phase
shift $\delta_0$ equals $\pi$ at $\epsilon=0$ for $T < 1.11 T_c$, in
accordance with the Levinson's theorem [which states that
$\delta=n\pi$ at $\epsilon=0$, where $n$ is the number of bound
states, except when an $S$-wave resonance occurs at zero energy, in
which case $\delta=(n+1/2)\pi$].  At temperatures below the
dissociation temperature $T_d=1.11T_c$ the phase shift decreases as a
function of energy, and is equal to $\pi/2$ at some energy
$\epsilon$. Such an occurrence of $\delta_0=\pi/2$ with
$d\delta(\epsilon)/d\epsilon <0$ does not represent a resonance; it is
an ``echo'' of the bound state which lies just below the continuum
\cite{Mcv67}.  At the location $\sin^2 \delta_0=1$ the scattering
cross section is enhanced when a bound state lies just below the
continuum, as was first noted by Wigner and discussed by Landau and
Lifshitz \cite{Lan58}.

At the $\Upsilon(1S)$ dissociation temperature $T_d=1.11T_c$,
$\delta_0=\pi/2$ at $\epsilon=0$, and the $L=0$ $\Upsilon(1S)$ and
$\eta_b(1S)$ resonances occur at $\epsilon=0$.  The phase shift
decreases fairly rapidly as a function of energy.  The energy
difference between the location of the maximum of $\sin^2 \delta_0$
and the half maximum is about 0.2 GeV, indicating a resonance with a
half width of about this magnitude.

For temperatures higher than $T_d$ the phase shift is zero at
$\epsilon=0$.  It increases rapidly with $\epsilon$ and reaches a
maximum before it decreases slowly.  One can associate the location
$\epsilon$ of the maximum phase shift in $\epsilon$ with the position
of the $\Upsilon(1S)$ and $\eta_b(1S)$ ``resonances''.  Strictly
speaking the enhancement at the phase shift maximum does not represent
a resonance, as the magnitude of the maximum phase shift is less than
$\pi/2$.  However when the transition temperature is slightly greater
than $T_d$, the magnitude of the maximum phase shift is close to
$\pi/2$, and it decreases continuously as the temperature increases.
The location $\epsilon$ of the maximum phase shift is a smooth
continuation of the bound state energy.  We can therefore speak of
these enhancements as ``resonances'', as they are the continuation of
the bound states into the positive energy domain.  We can choose to
identify the position of the maximum of the phase shift for $T >T_d$
with the location of the resonances as a function of temperature.  The
maximum value of $\sin^2\delta_0$ decreases as the temperature
increases.  The width of $\sin^2\delta_0$ increases, showing a broader
enhancement of the cross section as the temperature increases.  For
example, at $T=T_d+0.02 T_c$ the maximum occurs at $\epsilon=0.008$
GeV, and the separation between the maximum and the half maximum on
the higher energy side is about 0.30 GeV.  At $T=T_d+0.04 T_c$ the
maximum occurs at $\epsilon=0.018$ GeV, and the separation between the
maximum and the half maximum on the greater energy side is about 0.43
GeV.  The function $\sin^2\delta_0$ is not symmetrical with respect to
the maximum.  It initially rises rapidly, but then decreases slowly as
a function of $\epsilon$.  The width increases rapidly with
temperature.

We show in Fig.\ 5 the phase shift $\delta_1$ and $\sin^2 \delta_1$
for $L=1$ $Q$-$\bar Q$ scattering.  For $T$ slightly greater than the
phase transition temperature, the phase shift has a maximum at
$\epsilon\sim$ 0.1-0.2 GeV.  The maximum magnitude of the phase shift
is 0.54 at $\epsilon=0.12$ GeV for $T/T_c=1.01$, and is 0.39 at
$\epsilon=0.265$ GeV for $T/T_c=1.05$.  The maximum phase shift does
not reach $\pi/2$. The quantity $\sin^2\delta_1$ has a maximum value
of 0.22 at $T/T_c=1.01$, and a value of 0.14 at $T/T_c=1.05$.  These
enhancements are quite broad.  For example, for $T/T_c=1.01$, the
quantity $\sin^2\delta_1$ decreases slowly after it reaches the
maximum value, and the separation between the maximum and the half
maximum on the higher energy side is about 1.7 GeV.

\vspace*{1.8cm} \epsfxsize=300pt 
\includegraphics{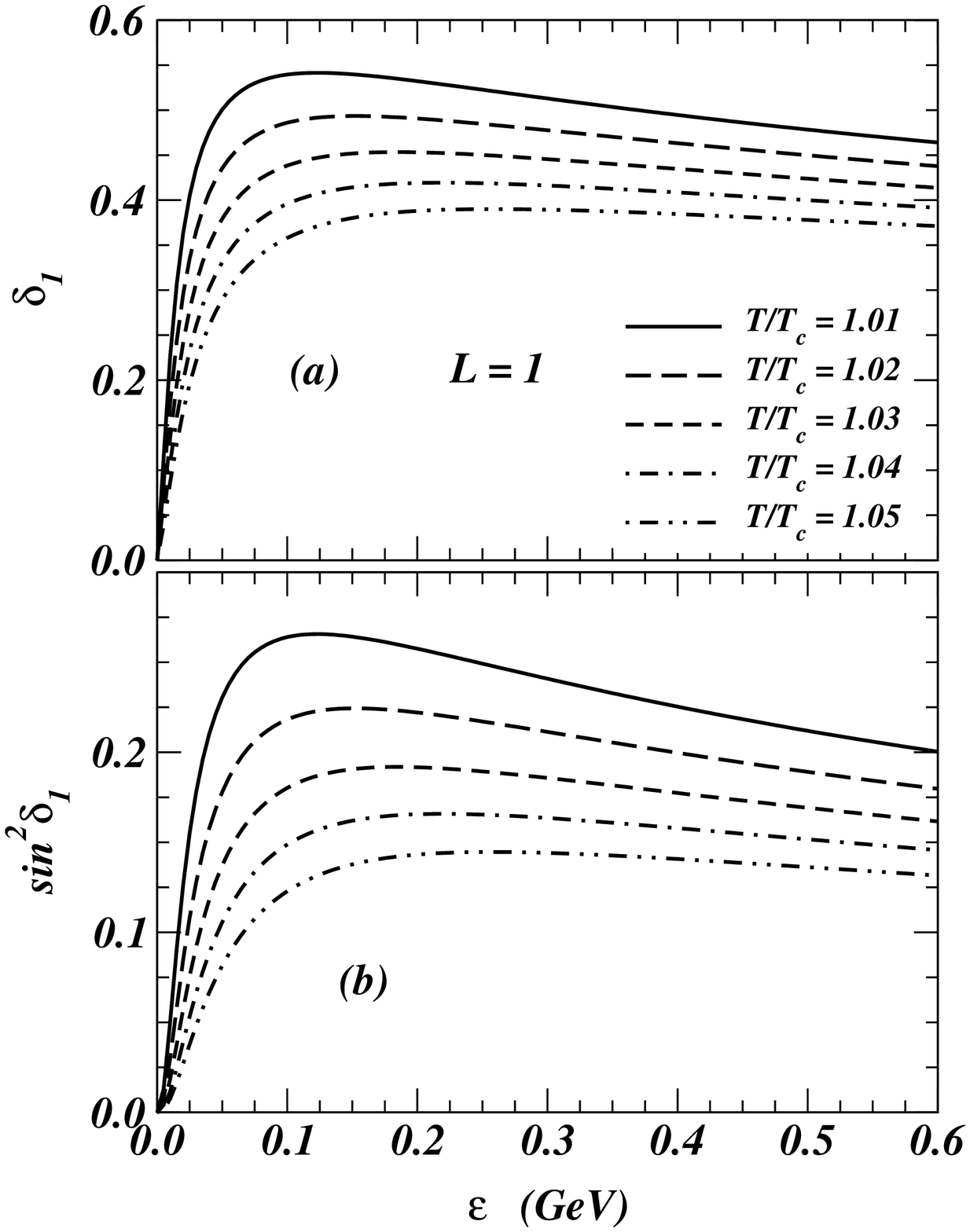}
\vspace*{+12.0cm}\hspace*{1cm}
\begin{minipage}[t]{12cm}
\noindent {{\bf Fig.\ 5 .} (a) The phase shift $\delta_1$ and 
{(b)}
$\sin ^2 \delta_1$ for the scattering of $Q$ and $\bar Q$ in the $L=1$
state as a function of the energy $\epsilon$ for various $T/T_c$.}
\end{minipage}
\vskip 4truemm
\noindent

\section{Cross Section for $\Upsilon(1S)$ + {\lowercase {\sl g}} $\to$
 {\lowercase {\sl b}} + $\bar {\lowercase {\sl b}}$ }

The quarkonia $\Upsilon(1S)$ and $\eta_b(1S)$ are stable in the
quark-gluon plasma, but have small binding energies (from 0 to about
0.0175 GeV).  They can dissociate into $b$ and $\bar b$ by collision
with quarks and gluons.

We shall focus our attention on the dissociation reaction
$\Upsilon(1S) +g \to b + \bar b $ through gluon absorption; the
dissociation of $\eta(1S)$ can be treated in a similar way.  For the
dissociation processes $\Upsilon (1S) +g \to b + \bar b$, the initial
quarkonium state is a $1S$ bound state, and the final state is a
$b\bar b $ state in the continuum.  This process is similar to the
photodissociation of a deuteron after photon absorption.  The
dissociation process can be studied using a multipole expansion of the
electromagnetic or chromodynamic fields, and the lowest multipole
orders are $M1$ and $E1$ absorption processes.  For low energy photons
or gluons, these processes can be treated in the long wavelength
limit, discussed in detail by Blatt and Weisskopf \cite{Bla52}.

Following Eq.\ (XII.4.33) of Blatt and Weisskopf \cite{Bla52}, the
$E1$ dissociation cross section for a transition from the initial
bound $1S$ state with energy $\epsilon_i=-B=-\gamma^2/m_Q$ to a final
$1P$ continuum state with energy $\epsilon=k^2/m_Q$, after absorbing a
gluon of energy $\epsilon-\epsilon_i$, is given by
\begin{eqnarray}
\label{eq:bla}
\sigma_{\rm dis}^{E1}=4 \times {\pi \over 3} 
\alpha_{\rm gQ} (k^2 +\gamma^2)k^{-1}I^2,
\end{eqnarray}
where the factor of 4 on the right hand side has been included for
color electric dipole contributions from both the $Q$ and the $\bar
Q$, (as distinct from the deuteron case, in which only the proton
contributes to the electric dipole matrix element).

The quantity $I$ of Eq.\ (\ref{eq:bla}) is the
radial overlap integral for the $E1$ electric dipole matrix element,
\begin{eqnarray}
\label{eq:I}
I=\int_0^{\infty} u_{1P}(r) u_{1S}(r) r dr .
\end{eqnarray}
where the wave function $u_{1P}(r)$ is given by Eqs.\
(\ref{eq:wf})-(\ref{eq:alp}). The $1S$ bound state wave function in
Eq.\ (\ref{eq:I}) is normalized according to
\begin{eqnarray}
\int_0^{\infty} |u_{1S}(r)|^2 dr =1.
\end{eqnarray}
The quantity $\alpha_{\rm
gQ}$ is an effective strong interaction coupling constant for a gluon
interacting with a heavy quark or antiquark.  Specifically, similar to 
Eq.\ (\ref{eq:8}) of the Appendix, we have 
\begin{eqnarray}
4 \pi  \alpha_{\rm gQ}
=|\langle f |g {\lambda^c \over 2} | i \rangle |^2 ,
\end{eqnarray}
where $|i\rangle$ is the initial state and $|f\rangle$ the final state
after the absorbing a gluon.  For our problem with a mixed-color heavy
quarkonium system in the QGP, we assume that $\lambda \gg \Delta$ in
Eq.\ (\ref{eq:eig0}) and so the initial and final states are
characterized by $a_0=1/\sqrt{2}=-a_8$. Then we obtain
\begin{eqnarray}
4 \pi  \alpha_{\rm gQ}
= {g^2 \over 6}. 
\end{eqnarray}
For these mixed-color states, if we represent the interaction between
$Q$ and $\bar Q$ by a gluon exchange, the associated coupling constant
will be $g^2[(-4/3)+(1/6)]/2$.  We identify this coupling constant
with the coupling constant extract from lattice calculations,
$|(-4/3)\alpha_{\rm eff}|$ of Eq.\ (\ref{eq:pot}).  Hence, the
coupling constant for the absorption of a gluon is then
\begin{eqnarray}
\alpha_{gQ}={1\over 6} \times {12\over 7} \times \left |- {4\over 3}\alpha_{\rm
eff} \right |.
\end{eqnarray}

Previously, the cross section for the dissociation of a color-singlet
heavy-quarkonium in collision with a gluon was obtained in the
short-distance approach by Peskin and Bhanot \cite{Pes79,Bha79} and
subsequently used by Kharzeev and Satz \cite{Kha94}, for the simple
case when the heavy-quarkonium can be treated as a hydrogenic system.
The calculation of Peskin and Bhanot uses the operator product
expansion, dispersion relations, and the sum of a large number of
diagrams in the large $N_c$ limit.  The leading order cross section
involves the square of the color electric field $\bbox{E}^2$, which
corresponds to the dissociation by an electric dipole radiation.  As
the physical process included in the evaluation of the dissociation
cross section in Peskin $et~al.$ \cite{Pes79,Bha79} and in Eq.\
(\ref{eq:bla}) of Blatt and Weisskopf \cite{Bla52} are the same, they
should give the same dissociation cross section.  We shall show the
equality of both results explicitly for the simple case considered by
Peskin and Bhanot in the Appendix.

The analytical result of Peskin and Bhanot is useful only for the
idealized case of a hydrogenic $1S$ initial color-singlet state and a
non-interacting $1P$ final state in the large $N_c$ limit.  The wave
functions of the heavy quarkonium systems of interest are not
hydrogenic.  The energy and the wave functions of the initial and
final states also depend sensitively on temperature.  The large $N_c$
limit is only a crude approximation for $N_c=3$.  We need to use the
more general results of Eq.\ ({\ref{eq:bla}) for our problem in which
both the initial and the final state wave functions have been obtained
numerically, without using the large $N_c$ limit.

\vspace*{2.0cm} \epsfxsize=300pt 
\includegraphics{qqbres.xdis.eps}
\vspace*{+11.3cm}\hspace*{1cm}
\begin{minipage}[t]{12cm}
\noindent { {\bf Fig.\ 6}. The $E1$ cross section for the dissociation
process  $\Upsilon(1S)+g \to b+\bar b$ after absorbing a gluon.}
\end{minipage}
\vskip 4truemm
\noindent

With the $L=1$ phase shift $\delta_1(r)$ and amplitude $\alpha_1(r)$
obtained in Section IV and the $L=0$ bound state wave function
obtained in Section III, the overlap integral (\ref{eq:I}) can be
evaluated to give the dissociation cross section of $\Upsilon(1S)$ by
gluon absorption.  We show in Fig.\ 6 the dissociation cross section
of $\Upsilon(1S)$ by gluon absorption as a function of the final
continuum state energy $\epsilon$.  The cross section is zero at
$\epsilon=0$, rises to a maximum and then decreases with increasing
$\epsilon$.  The dependence of the $E1$ $\Upsilon(1S)$ dissociation
cross section on energy is similar to that of the $E1$ deuteron
photodissociation cross section as shown in Fig.\ XII.4.1 of
\cite{Bla52}.

Because the mass of the $b$ quark is large, the $M1$ cross section is
small relative to the $E1$ cross section, and can be neglected.

It should be noted that the dissociation cross section has been
calculated for the exclusive channel $g+\Upsilon(1S)\to b +\bar b$.
As the energy increases, the cross section for this specific channel
decreases (Fig.\ 6) and  other reaction channels such as
$g+\Upsilon(1S)\to b +\bar b + n g + n' (q+\bar q)$ will make
important contributions to the dissociation of the $\Upsilon(1S)$.  It
will be of great interest to evaluate these cross sections in future work.

\section{Discussion}

We have studied the dissociation of a heavy quarkonium state in a
quark-gluon plasma using the finite-temperature potential $V_{Q\bar
Q}(r)$ inferred from lattice gauge calculations of Karsch $et~al.$
\cite{Kar00} Focusing our attention on states that can be directly or
indirectly detected by dilepton decay, we can classify quarkonia into
three groups: (1) quarkonia which dissociate spontaneously in hadronic
matter: $\psi'$, $\chi_c$, $J/\psi$, $\Upsilon''$, $\chi_b$,
$\Upsilon'$; (2) quarkonia which are stable in hadronic matter but
dissociate spontaneously in the quark-gluon plasma: $\chi_b$; and (3)
quarkonia which are stable in hadronic matter and the quark-gluon
plasma below a dissociation temperature: $\Upsilon(1S)$ with the
dissociation temperature of 1.11$T_c$.

From this classification, one notes that the $\chi_{b1}$ and
$\chi_{b2}$ mesons are good indicators of a quark-gluon plasma, as
they are stable in hadronic matter but dissociate spontaneously in a
quark-gluon plasma.  The $\Upsilon(1S)$ (and $\eta_b(1S)$) have the
distinction that they are the only stable quarkonia in the quark-gluon
plasma, although the binding energies are rather small.

For a quark-gluon plasma at a temperature below the $\Upsilon(1S)$
dissociation temperature $T_d$, a direct observation of the position
of the $\Upsilon(1S)$ using decays to dileptons is of great interest.
The mass of the $\Upsilon(1S)$ state in the quark-gluon plasma is
$2m_b+\epsilon(1S)$, where $\epsilon(1S)$ lies between -0.0175 GeV and
0 GeV (Fig.\ 3), and the current $b$ quark mass $m_b$ is $4.2 \pm 0.2$
GeV \cite{pdg00}.  The invariant mass of the dilepton peak from the
decay of the $\Upsilon(1S)$ in the quark-gluon plasma will be
substantially lower than the dilepton invariant mass of 9.46 GeV seen
in the decay of the $\Upsilon(1S)$ in free space.  The fraction of
dileptons from the decay of the $\Upsilon(1S)$ inside the plasma
depends on the lifetime of the quark-gluon plasma and the dilepton
decay partial width of $\Upsilon$ in the plasma.  This fraction may be
rather small, making the observation of the dilepton energy peak shift
a difficult task.

We have examined the dissociation cross section for the exclusive
channel $g+\Upsilon(1S) \to b + \bar b$ through a color $E1$
transition.  We have also studied the phase-shifts for $Q$-$\bar Q$
collisions in the continuum. We find that there are $\Upsilon(1S)$ and
$\eta_b(1S)$ resonances at $\epsilon=0$ for $L=0$ at the dissociation
temperature $T_d=1.11T_c$, with a half width of $\Gamma/2 \sim 0.2$
GeV.  The occurrence of these $L=0$ resonances just above $\epsilon=0$
is a characteristic of the quark-gluon plasma, as they are not
expected from other sources.  These resonances will decay in the
quark-gluon plasma and appear as separated $b$ and $\bar b$ quarks.
After the quark-gluon plasma cools to become hadronic matter, the $b$
and $\bar b$ quarks will hadronize into $B$ and $\bar B$ mesons.  The
invariant mass of this pair will show a resonance at low relative
kinetic energies.  However, direct detection of $B$ and $\bar B$
mesons is a difficult task whose prospects may improve with future
advances in detector technology.

The author would like to thank Drs. M. Creutz, Keh-Fei Liu,
L. McLerran, and S. Ohta for valuable discussions on lattice gauge
theory.  The author also wishes to thank Drs. T. Barnes, V. Cianciolo,
H. Crater, and G. Young for helpful discussions.  This research was
supported by the Division of Nuclear Physics, Department of Energy and
the Laboratory Directed Research and Development Program of Oak Ridge
National Laboratory under Contract No. DE-AC05-00OR22725 managed by
UT-Battelle, LLC.

\vskip 0.5cm
\appendix

{\bf Appendix : Approximate gluon dissociation cross section for a
color-singlet heavy quarkonium}

We would like to show how Eqs.\ ({\ref{eq:bla})-({\ref{eq:I}) lead to
the result of Peskin and Bhanot \cite{Pes79,Bha79} for the special
case they considered. They assumed that the initial ground state of the
color-singlet heavy quarkonium can be approximately described by a
hydrogenic $1S$ wave function
\begin{eqnarray}
u_{1S}(r)=2 \gamma^{3/2} r e^{-\gamma r},
\end{eqnarray}
where $\gamma^2=2\mu_{Q\bar Q}B$, $B=\mu_{Q\bar Q}\alpha_{\rm
singlet}^2/2$, and $\gamma=|\mu_{Q\bar Q}\alpha_{\rm singlet}|$.  Here,
$\alpha_{\rm singlet}$ is the color-singlet coupling constant,
relating to the strong interaction coupling constant $\alpha_s$ by
\begin{eqnarray}
\label{eq:5}
\alpha_{\rm singlet}=-{N_c^2-1 \over 2 N_c} \alpha_s,
\end{eqnarray}
where $N_c=3$ is the number of colors.  Peskin and Bhanot used the
large $N_c$ approximation and the above color-singlet effective
coupling constant for $N_c=3$ is approximated in this large $N_c$
limit by
\begin{eqnarray}
 \lim_{\rm large~N_c} ~\alpha_{\rm singlet} =
-{N_c \over 2 }\alpha_s=- {3 \over 2 }\alpha_s.
\end{eqnarray}
Peskin and Bhanot considered the final color-octet state to be
approximated by the $1P$ wave function as in the free-particle case.
The corresponding wave function is
\begin{eqnarray}
u_{1P}(r)={\sin kr \over kr} - \cos kr.
\end{eqnarray}
The overlap integral $I$ of Eq.\ (\ref{eq:I}) then gives
\begin{eqnarray}
I&=&\int_0^{\infty} u_{1P}(r) u_{1S}(r) r dr \nonumber\\
&=&{ 16 \gamma^ {5/2} k^2 \over (k^2 +\gamma^2)^3 }.
\end{eqnarray}
Using $k^2=2\mu_{Q\bar Q}\epsilon$ and the relation between $\gamma$
and the binding energy $B$, Eq.\ (\ref{eq:bla}) from Blatt and
Weisskopf gives the dissociation cross section
\begin{eqnarray}
\label{eq:dis}
\sigma_{\rm dis}^{E1}((Q\bar Q)_{1S}+g \to Q+\bar Q)= { 4 \pi \over 3}
(32)^2 { \alpha_{\rm gQ} \over \alpha_{\rm singlet}^2 m_Q^2}
{(\epsilon/B)^{3/2} \over (\epsilon/B+1)^5 }.
\end{eqnarray}
The quantity $\alpha_{\rm gQ}$ in the above equation is the effective
coupling constant for the interaction of a gluon with the $Q$ (or
$\bar Q$) of the quarkonium.  In the problem studied by Peskin
$et~al.$ \cite{Pes79,Bha79}, the initial $1S$ state is in the
color-singlet state
\begin{eqnarray}
\label{eq:6}
|1 \rangle = {1\over \sqrt{3}} \sum_{ij} | 3 i, \bar 3 j\rangle,
\end{eqnarray}
and the final $1P$ state is in a color-octet state with color component $c$
\begin{eqnarray}
| 8\, c \rangle = {1\over \sqrt{2}} \sum_{ij} \lambda_{ij}^c | 3 i, \bar 3
  j\rangle.
\end{eqnarray}
The fermion-(gauge boson) vertex in a Feynman diagram is associated
with $-ie\gamma^\mu$ in QED and with $g\gamma^\mu{\lambda}^c/2$ in
QCD where $c$ is the color index of the gluon.  Thus, in going from
the QED case of Blatt and Weisskopf to the QCD case of Peskin and
Bhanot, $e$ in QED is replaced by $g{\lambda^c}/2$ in QCD.  The
value of $e^2 =4 \pi \alpha_{\rm QED}$ in QED of Blatt and Weisskopf
is replaced in QCD by
\begin{eqnarray}
\label{eq:8}
|\langle 8c |g {\lambda^c \over 2} | 1 \rangle |^2 
=4 \pi \alpha_{\rm gQ}
\end{eqnarray}
where $\lambda^c$ acts only on the $Q$ (or $\bar Q$).  Eqs.\
(\ref{eq:6})-(\ref{eq:8}) gives
\begin{eqnarray}
\label{eq:9}
\alpha_{\rm gQ}
= {g^2 \over 4\pi \times  6}={\alpha_s \over 6}.
\end{eqnarray}
The above equations lead to
\begin{eqnarray}
\label{eq:pes}
\sigma((Q\bar Q)_{1S} + g \to Q + \bar Q) = {2 \over 3} \pi \left (
{32 \over 3}\right )^2 \biggl |\left ( {16 \pi \over 3 g^2 }\right )
{1 \over m_Q^2} \biggr | 
~{ (\epsilon/B)^{3/2} \over (\epsilon/B+1)^5},
\end{eqnarray}
which is the same as Eq.\ (4.4) of Bhanot and Peskin
(Ref. \cite{Bha79}), demonstrating the equivalence of the descriptions
of Blatt and Weisskopf and Peskin and Bhanot.

It should be pointed out that the large $N_c$ limit is only a crude
approximation for $N_c=3$.  If one does not invoke the large $N_c$
limit, we then have for $N_c=3$, $\alpha_{\rm
singlet}=-(4/3)\alpha_s$.  Hence, we obtain a more accurate
dissociation cross section from Eqs.\ (\ref{eq:dis})-(\ref{eq:9})
\begin{eqnarray}
\label{eq:new}
\sigma((Q\bar Q)_{1S} + g \to Q + \bar Q) =\left ( {9 \over 8}\right
)^2 {2 \over 3} \pi \left ( {32 \over 3} \right )^2 \biggl |\left ({16
\pi \over 3 g^2}\right ) {1 \over m_Q^2}  \biggr | ~{ (\epsilon/B)^{3/2}
\over (\epsilon/B+1)^5},
\end{eqnarray}
which is $(9/8)^2$ times the approximate
large-$N_c$ limit result of Peskin and Bhanot \cite{Pes79,Bha79}.

One notes that from this cross section expression (\ref{eq:new}), the
peak of the dissociation cross section is located at $\epsilon=3B/7$
and the maximum dissociation cross section from Eq.\ (\ref{eq:new}) is
\begin{eqnarray}
\sigma_{\rm max}( (Q\bar Q)_{1S}+g \to Q + \bar Q) = {18.90\over \alpha_s
m_Q^2}.
\end{eqnarray}
If one uses the hydrogenic description for $J/\psi$ with
$\alpha_s=0.3$, then $B=0.06$ GeV and the maximum of the gluon
dissociation cross section is located at $\epsilon=0.025$ GeV with a
maximum of 10.90 mb.  For a color-singlet $\Upsilon$ with
$\alpha_s=0.2$, then $B=0.089$ GeV and the maximum cross section is
located at 0.038 GeV with a maximum of 1.5 mb.

\end{document}